# Forced ignition and oscillating flame propagation in fine ethanol sprays

Qiang Li[a], Huangwei Zhang[a, *]

[a] *Department of Mechanical Engineering, National University of Singapore, 9 Engineering Drive 1, Singapore 117576, Republic of Singapore*

______________________________________________________________________

**Abstract**

The present work investigates forced ignition and oscillating propagation of spray flame in a mixture of fine ethanol droplets and air. Eulerian-Eulerian method with two-way coupling is used and detailed chemical mechanism is considered. Different droplet diameters and liquid fuel equivalence ratios (ER) are studied. The evaporation completion front (ECF) is defined to study the interactions between the evaporation zone and flame front. The gas composition at the flame front is quantified through an effective ER. The results show that the kernel trajectory is considerably affected by droplet size and liquid ER. Generally, the flame ER reaches the maximum when the ECF start to move from the spherical center. It gradually decreases and reaches a constant value when the flame freely propagates. Quasi-stationary spherical flame is observed when the liquid ER is low, whilst kernel extinction/re-ignition appears when liquid ER is high. These flame behaviors are essentially affected by the heat conduction and species diffusion timescales between the droplet evaporation zone and flame front. The dependence of the minimum ignition energy (MIE) on liquid ER is U-shaped and there is an optimal liquid equivalence ratio range ($ER_o$) with the smallest MIE. Long and short ignition failure modes are observed, respectively for small and large liquid ERs. When liquid ER is less than $ER_o$ (long failure mode), larger energy is required to initiate the kernel due to very lean composition near the spark. For liquid ER larger than $ER_o$ (short mode), ignition failure is caused by strong evaporative heat loss and rich gas composition due to heavy droplet loading. Flame speed oscillation is seen when initial droplet size is above 10 μm. The oscillation frequency linearly increases with the liquid ER. Meanwhile, larger droplets lead to smaller oscillation frequencies.

______________________________________________________________________

*Corresponding author. Email address:

huangwei.zhang@nus.edu.sg (H. Zhang)

## 1. Introduction

Sprayed liquid fuels are widely used in advanced propulsion system, e.g., aero-engine, rocket engine, and gas turbine. Upon atomization and breakup, the fuel droplets are dispersed in the combustor, and interact with the propagating flame front, through mass, momentum, and heat transfer. The droplet properties and dynamic behaviors would strongly influence the unsteady flame dynamics [1,2]. Therefore, it is of great significance to understand the effects of droplet properties on unsteady flame behaviors, such as ignition and propagation.

Compared to gaseous flames, the existence of dispersed liquid fuel droplets may bring intriguing features in the ignition process. Comprehensive experiments on spark ignition of different fuel mists by Ballal and Lefebvre [3] show that the initial droplet size has considerable effects on the minimum ignition energy (MIE). It is also found that for pure fuel mists, adequate concentration of fuel vapor near the spark is required for flame initiation. Besides, small droplets are beneficial for spray flame ignition due to fast heating and evaporation [4]. Meanwhile, the composition of heterogeneous mixtures is also critical for successful ignition [5,6]. For ignition failure, based on their different timescales, two modes are defined, i.e., short mode and long mode [1]. Short mode of ignition failure is also observed in direct numerical simulations, characterized by a low-temperature flame kernel [5]. Nonetheless, fuel vapour starvation induced by slow evaporation speed leads to long mode failure [7]. The two modes are also observed in experiments and evaluated by a reference timescale of the flame overdrive [8]. More recently, Li et al. [9] investigate the transient evolution of evaporation zone and its relative location with the flame front in the ignition of spray flame. The results indicate that failed and critical successful ignition events occur only in purely gaseous mixture near the flame front.

For spray flame propagation, interactions between the flame front and evaporating droplets have also been studied. Previous studies show that the droplet size and loading affect their distributions relative to the flame front [10–12]. Large droplets may penetrate the post-flame zone and continue evaporating there, while small ones can fully vaporize in the pre-flame zone. Atzler et al. [13] observed flame speed oscillation in their experiments on iso-octane spray flames for overall lean mixture and large droplet size. One proposed explanation for this phenomenon is that the oscillation is caused by the velocity difference between droplet and gas in the evaporation zone near the flame front. A similar oscillation phenomenon is also reported for ethanol spray flame [14]. Meanwhile, it is found that the interactions between vaporizing droplets and flame front play an important role in determining the flame speed oscillation [13,15]. The evaporation droplets periodically penetrate the post-flame zone due to their large inertia when the gaseous flame oscillates. This shift of the evaporation zone brings significant variation for the equivalence ratio at the flame front.

However, interplay between the evolving flame front and evaporating droplets and how this affects the spray flame dynamics have not been fully understood, due to the strong unsteadiness and multi-faceted gas-liquid coupling. In this work, detailed numerical simulations will be performed on forced ignition and propagation of ethanol spray flame based on Eulerian-Eulerian approach. Evolutions of the droplet evaporation zone relative to the travelling flame front will be studied in our simulations, through quantifying the movement of the droplet evaporation front, which is a two-phase contact surface separating the gas-only and droplet-carrying mixtures. The objectives of this work are to clarify the effects of the fuel droplet and flame interactions on the various flame development stages, which include: (1) flame kernel initiation and expansion; (2) short and long ignition failure in fuel sprays; and (3) oscillating flame propagation in fine ethanol sprays.

## 2. Physical and numerical models

### 2.1 Physical model

Forced flame ignition in ethanol sprays is numerically studied with a one-dimensional spherical configuration. The computational domain is $0 \leq r \leq r_W$, where $r$ is the radial coordinate and $r_W$ is the domain radius, taken as 20.48 cm, unless otherwise stated. Initially, a mixture of ethanol droplets and air is uniformly distributed in the domain. The ethanol sprays are dilute ($< 0.1\%$ by vol.), and the droplets spherical and mono-sized. The liquid fuel sprays are characterized by initial droplet diameter, $d_0$, and liquid equivalence ratio (ER), $\phi_{l,0}$. The latter is the mass ratio of liquid fuels to air, scaled by the counterpart ratio under stoichiometry. The droplet diameter, $d_0$, varies from 5 to 15 $\mu$m. The considered $\phi_{l,0}$ ranges from 1.4 to 6.0. The rich composition mimics preferential accumulation of sprayed droplets near the spark; based on our simulations the sprays with $\phi_{l,0} < 1.4$ are not ignitable even if very large ignition energy (e.g., > 6 mJ) is deposited. Moreover, before ignition, both air and ethanol droplets are at rest, with a temperature of 298 K. The initial gas pressure is 1 atm.

According to previous studies [8–10,16], the interactions between the evaporating droplets and reaction fronts significantly modulate gaseous flame dynamics. As in our recent theoretical work [9,16], we introduce a concept of evaporation completion front (ECF), $R_c$, where the droplets are just completely vaporized (see Fig. 1, identified as the location where the droplet diameter is critically less than 0.1 μm). The clear-cut boundary at which the droplets are fully evaporated is observed through laser sheet imaging of laminar spray flames, e.g., in [13][15]. Besides, the flame front (FF), $R_f$, is the location with maximum heat release rate (HRR). If their relative distance $\Delta R = R_c - R_f > 0$, then the ECF lies at the unburned zone, indicating that the mixture at the FF is gaseous, i.e., fuel vapor and air (droplets already fully gasified). This is termed as a homogeneous (HM) spray flame (see the schematic in Fig. 1a). If $\Delta R < 0$, the ECF is behind the FF. Hence, the mixture at the FF is composed of evaporating droplets, vapor, and air. This is a heterogeneous (HT) flame (see Fig. 1b).

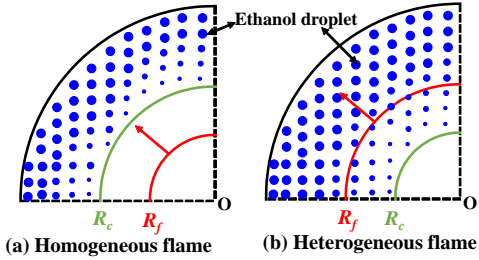

(a) Homogeneous flame     (b) Heterogeneous flame

Fig. 1. (a) Homogeneous and (b) heterogeneous spray flames. $R_f$: flame front; $R_c$: evaporation completion front.

To quantify the evolving gas composition due to evaporating sprays, an element-based method [17] is used to calculate the effective ER of the gas phase

$$\phi_e = [3(\tfrac{X_C}{2} + \tfrac{X_H}{6})]/[X_O - \tfrac{1}{2}(\tfrac{X_C}{2} + \tfrac{X_H}{6})], \quad (1)$$

where $X$ denotes the mole fraction of an element (i.e., C, H or O). Although Eq. (1) is general, we always calculate $\phi_e$ at the FF, $\phi_{e,f}$, to assist our discussion. We term it as *flame ER* hereafter.

*2.2 Numerical implementation*

An in-house code, A-SURF [18], is used to simulate the forced ignition in ethanol sprays. A Eulerian−Eulerian method is employed to describe the gas−liquid two-phase flows, considering two-way coupling between the continuous phase and dispersed phase for mass, momentum, and energy exchanges.

The droplet evaporation rate is estimated as $\dot{m} = \pi d \rho_s D_{ab} Sh \ln(1 + B_M)$ [19], where $d$ is the droplet diameter, $\rho_s$ the vapour at the droplet surface, and $D_{ab}$ vapor mass diffusivity in the gas mixture. $B_M$ is the Spalding mass transfer number, and $Sh$ is the Sherwood number. Only Stokes drag force is considered for droplet momentum equation, which is modelled using Schiller and Naumann's correlation, i.e., $F_s = m_d/\tau_r \cdot (u_g - u_d)$. $m_d$ is the mass of a single droplet, and $\tau_r$ is the droplet momentum relaxation time. $u_g$ and $u_d$ are the gas and droplet velocities, respectively. Besides, evolution of the droplet temperature is governed by convective heat transfer, $h_c A_d (T_g - T_d)$, and latent heat absorption, $\dot{m}L_v$. $h_c$ is the convective heat transfer coefficient, and $A_d$ is the droplet surface area. $T_g$ and $T_d$ are the gas and droplet temperatures, respectively. $L_v$ is the latent heat of vaporization. For interested readers, please refer to the detailed formulations in the supplementary document.

Zero gradient conditions for gas phase species, temperature, and liquid phase quantities (diameter, temperature, number density) are enforced at the spherical center ($r = 0$) and right boundary ($r = r_W$). Zero velocities for gas and droplets are assumed for both boundaries. To model spark ignition, an source term, $q_{ig}(r,t)$, is added for the energy equation of the gas phase [20]

$$q_{ig}(r,t) = \begin{cases} \frac{E_{ig}}{\pi^{1.5} r_{ig}^3 t_{ig}} exp\left[-\left(\frac{r}{r_{ig}}\right)^2\right], & t < t_{ig} \\ 0, & t \geq t_{ig} \end{cases} \quad (2)$$

where $t$ is time and $E_{ig}$ is the ignition energy. The spark size $r_{ig}$ and duration $t_{ig}$ are assumed to 400 μm and 400 μs, respectively [21].

Adaptive mesh refinement (up to eight levels, leads to finest cell size of 16 μm) is used, to accurately resolve the flame and evaporation fronts. The time integration, diffusive flux, convective flux, and droplet quantities are calculated by Euler explicit method, central differencing, MUSCL-Hancock, and upwind schemes, respectively. A detailed ethanol mechanism (57 species, 383 reactions) [22] is used. The chemistry is integrated using the point implicit method, and detailed transport is calculated with CHEMKIN and TRANSPORT packages [23]. The accuracies of the A-SURF solver for gaseous and spray flames are already validated, e.g., in Refs. [18,24].

**3. Results and discussion**

*3.1 Flame kernel development*

Three cases are selected to discuss flame kernel development in spark ignition of ethanol sprays. Their initial conditions are: (a) case 1: $d_0 = 5$ μm, $\phi_{l,0} = 1.75$; (b) case 2: $d_0 = 5$ μm, $\phi_{l,0} = 3.0$; and (c) case 3: $d_0 = 15$ μm, $\phi_{l,0} = 3.0$. The flames are initiated with the same ignition energy $E_{ig} = 6$ mJ. Figure 2 shows the trajectories of FF and ECF and the fuel vapor mass fraction distribution in the $r$-$t$ diagram. The evolutions of the corresponding front distance $\Delta R$ and flame ER $\phi_{e,f}$ are provided in Fig. 3.

For case 1, the igniting kernel is established and quickly expands before $t > 2.39 t_{ig}$ (point A in Fig. 2a). During this period, as seen from Fig. 3(a), the evaporating droplets are dispersed in the entire burned area. This leads to continuously increased flame ER, shown in Fig. 3(b). Afterwards, the droplets at the spherical center have been critically fully gasified due to their relatively long exposure to the hot spark, and subsequently the ECF starts to move outwardly with the FF. This is defined as the deviation point between the ECF and spherical center (same for case 2 and 3), marked as DP in Fig. 3(a) and the flame ER reaches the maximum of around 2.4 for case 1. When $R_f = 1.59\,r_{ig}$, the ECF overtakes the propagating FF, characterized by zero-crossing of the $\Delta R$ curve. This signifies a transition from HT ($\Delta R < 0$) to HM ($\Delta R > 0$) spray flames. The flame ER then monotonically decreases towards 0.45 at point A, slightly lower than the lower flammability limit (LFL) of ethanol (~0.5 [25]).

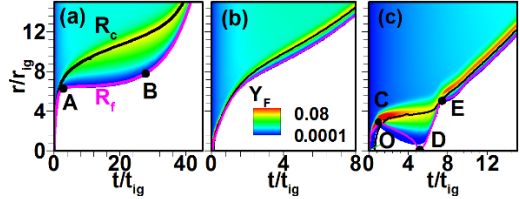

Fig. 2. $r$-$t$ diagrams of FF and ECF trajectories and fuel vapor mass fraction in (a) case 1, (b) case 2, and (c) case 3. $E_{ig}$ = 6 mJ.

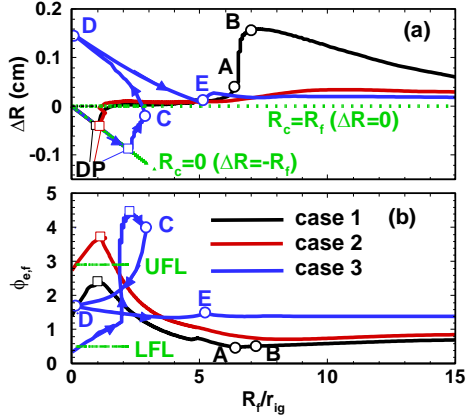

Fig. 3. Evolutions of (a) $\Delta R$ and (b) $\phi_{e,f}$ in cases 1-3. LFL/UFL: lower/upper flammability limit.

Beyond point A, the trajectories of ECF and FF are bifurcated, and $\Delta R$ increases quickly (see Fig. 3a). It is shown in Fig. 2(a) that the FF becomes quasi-stationary between $2.39 t_{ig}$ and $24 t_{ig}$ (A→B), with limited change of flame radius. Based on our results (not shown), the HRR of this period decreases significantly, resulting from insufficient fuel vapor at the FF, which is because the evaporation zone moves well ahead of the FF (large $\Delta R$). This is evident in Fig. 2(a) for the $Y_F$ contour between the two instants. The quasi-stationary spherical flame is intensified again through the continuous fuel vapor supply because the front distance $\Delta R$ gradually deceases after B, as shown in Fig. 3(a). Therefore, expansion of the spray flame is resumed at B. Subsequently, two fronts move synchronously, and the flame ER is almost constant (about 0.65, 37% $\phi_{l,0}$).

Conversely, for case 2 in Fig. 2(b), due to increased liquid ER ($\phi_{l,0}$=3.0), the fuel vapour from ECF to the FF is generally more sufficient than that in case 1. This can be confirmed by higher flame ER (about 3.8) at point DP in Fig. 3(b). Therefore, transient stationary flame is not observed; instead, two fronts, as a complex, are always coupled in a HM flame when they propagate in a self-sustaining fashion. Consequently, its evolutions of $\Delta R$ is different from that of case 1 after $R_f = 6 r_{ig}$ (see Fig. 3a).

Case 3 in Fig. 2(c) exhibits more interesting evolutions of the ECF and FF. Four stages are identified with several milestones (O, C, D and E). Specifically, stage 1 (O→C, $t \leq t_{ig}$) is the sparking period. Due to relatively large droplets (15 $\mu$m) and liquid ER (3.0), a HT flame is observed and the entire burned zone is full of evaporating droplets (i.e., $R_c = 0$) until $R_f = 2.32 r_{ig}$ (DP in Fig. 3a), much larger than those in cases 1 and 2. Near point DP, considerable fuel vapor is available at the FF due to continuous evaporation from the post-flame droplets and the local flame ER quickly rises to over 4.4, much higher than the upper flammability limit (UFL) of 2.9 [25]. Even with such rich composition, the FF still gradually moves outwardly and reaches a largest radius at point C, as seen from Fig. 2.

The spark is off at point C and stage 2 (C→D) signifies kernel decaying. The kernel has not reached self-sustaining state because of the strong heat loss by the heavy droplet loading and very rich composition near the kernel as discussed above. Due to termination of the external energy deposition, the flame kernel is gradually extinguished. Meanwhile, the two fronts are still approaching ($\Delta R \downarrow$), and immediately after C the ECF catches up with the decaying FF, leading to a transition from HT to HM flames. Due to continuously reduced heat supply from the FF, the movement of the ECF ahead of the flame slows down from C to D (Fig. 2c), but with a quickly lengthened front distance $\Delta R$ (Fig. 3a) due to the receding flame. Meanwhile, from Fig. 3(b), the flame ER decreases precipitously from 4.0 at C to 1.7 at D. Nonetheless, after D, the front distance $\Delta R$ is gradually reduced, as displayed in Fig. 3(a), which means that the two fronts come closer (but still as a HM flame). The flame ER is close to 1.38. D→E, as stage 3, is flame re-ignition process, and D is deemed the re-ignition point. After point E (stage 4), the FF and ECF are coupled as an outwardly expanding HM flame, featured by steady front distance and flame ER (about 1.38, 46% $\phi_{l,0}$) in Fig. 3.

To further understand the quasi-stationary spherical flame and re-ignition transients discussed above, the timescales behind them are estimated, including chemical ($\tau_r$), fuel vapor diffusion ($\tau_m$), thermal ($\tau_t$), and evaporation ($\tau_e$) timescales. $\tau_r$ is calculated from $\tau_r = \delta/S_{u,0}$. $\delta$ is the flame thickness, whilst $S_{u,0}$ is the laminar flame speed. $\tau_m$ is from $\tau_m = l_m^2/\overline{D}_{ab}$, whereas $\tau_t = l_t^2/\overline{\alpha}$. $\overline{D}_{ab}$ and $\overline{\alpha}$ are the ethanol mass diffusion and thermal diffusion coefficients. $l_m$ and $l_t$ is the mass and thermal diffusion length. Their definitions are also provided in the supplementary document. $\tau_e$ is estimated from $\tau_e = m_0/\dot{m}_{max}$. $m_0$ is the initial mass of a droplet, and $\dot{m}_{max}$ is the maximum evaporation rate in evaporation zone. The evolutions of the foregoing timescales in cases 1-3 are shown in Fig. 4.

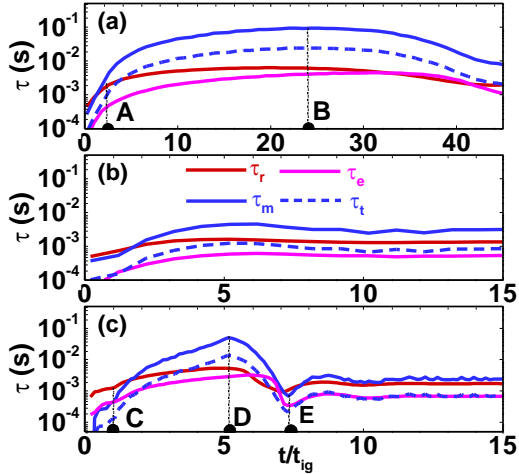

Fig. 4. Evolutions of different timescales in (a) case 1, (b) case 2, and (c) cae 3. A-E: milestones in Figs. 2 and 3.

We first look at case 2 in Fig. 4(b), in which the steady ECF/FF coupling is observed. Here $\tau_r$ ($\tau_t$) is close to $\tau_m$ ($\tau_e$). Heat and fuel species diffusion timescales between two fronts are comparable. In case 1, four timescales increase in the stage of quasi-stationary spherical flame. Near A, $\tau_m > \tau_r$, while $\tau_t > \tau_e$. This indicates that the fuel vapor diffusion from the evaporating zone to the reaction zone becomes too slow to maintain a self-sustaining FF. Meanwhile, the heat diffusion also limits the droplet heating. Therefore, decelerated fuel species and heat diffusion between the ECF and FF result in transient propagation termination. Closer inspection of Fig. 4(a) shows that, after B, $\tau_m$ and $\tau_t$, as the limiting factors, show a slightly earlier reduction compared to the rest two. This means that the variations of heat and species diffusion timescales precede the changes of the flame reactivity and droplet evaporation. Indeed, they are sequentially followed by reduction of $\tau_r$ and then $\tau_e$, as clearly demonstrated in Fig. 4(a). Therefore, the slow accumulation of the fuel vapor towards the FF makes the quasi-stationary spherical flame expand again.

For case 3 in Fig. 4(c), after C, $\tau_r$ gradually increases, and therefore the chemical reaction slows down because of the weakening flame kernel. Both $\tau_m$ and $\tau_t$ increase and peak at D, due to the largest diffusion length there. The flame is re-ignited after D and the four timescales continuously decrease. The four timescales peak sequentially, like case 1 in Fig. 4(a). This indicates the fuel accumulation near the kernel is significant to induce the re-ignition at E, which can also be confirmed from the fuel vapor mass fraction distribution between $R_c$ and $R_f$ in Fig. 2(c). This, in turn, intensifies the flame and hence droplet heating. After E, as an outwardly propagating spray flame, the corresponding timescales of heat and species diffusion are close again.

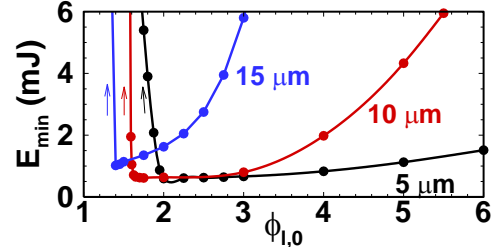

Fig. 5. Change of minimum ignition energy with liquid equivalence ratio for different droplet diameter.

### 3.2 MIE and ignition failure mode

The minimum energies for igniting ethanol sprays of different liquid ERs are presented in Fig. 5. The MIE is determined by trial-and-error method with error < 2%. Evident from Fig. 5 is that the MIE-$\phi_{l,0}$ curves are $U$-shaped, for all three studied droplet sizes. They can be approximately divided into left and right branches. Meanwhile, there exists a liquid ER range, ER$_o$, within which the smallest MIEs are reached and/or MIE has a weak dependence on $\phi_{l,0}$. For instance, for $d_0 = 5~\mu m$, ER$_o$ is 2.0–2.5. This range is shifted leftward with increased $d_0$.

For the left branch, which generally has the liquid ER close to unity (specifically, 1.0-2.0), the MIE increases precipitously if the liquid ER is slightly reduced. In our simulations, the mixtures with liquid ERs of this range become not ignitable even with very high ignition energy (above 6 mJ). We take case 1 as an example to justify the above peculiar phenomenon. As seen from Fig. 3(b), the flame ER of a developing kernel (such as at point A) is lower than LFL. Successful propagation relies on the fuel vapor diffusion from ECF to FF for the fuel-lean condition. Smaller liquid ER corresponds to less liquid droplets in the evaporation zone and therefore lower flame ER. The kernel needs more external energy to obtain sufficient fuel vapor. Hence, slightly decreased liquid ERs (hence less fuel vapor) should be offset by larger ignition energy to gasify more fuel droplets for the continued growth of the embryonic kernel, which necessitates quickly increased ignition energy.

For the right branch, due to high liquid ER, the transient fuel vapor concentration near the kernel can be above the UFL (for instance, the instantaneous flame ER can be above 4.0 in case 2, in Fig. 3b), although it would ultimately fall in the flammable range. This indicates that droplet evaporation heat loss of the flame kernel becomes dominant. Increased liquid ER leads to increased fuel vapor concentration in the kernel and higher evaporative heat loss. Hence, the MIE increases with increased liquid ER on the right branch.

From our simulations, long and short ignition failure modes, as discussed by Mastorakos [1], are observed. In Fig. 6, we show the examples of two modes, through the $r$-$t$ diagram of OH mass fraction for $\phi_{l,0}$=1.75 and 6.0, overlaid by the ECF and FF trajectories. The time history of flame ER in the ignition process is also provided. Here $E_{ig}$= 1.0 mJ and $d_0$= 5 μm. This $E_{ig}$ counts for 19% and 66% of the MIEs for $\phi_{l,0}$=1.75 and 6.0 cases, respectively.

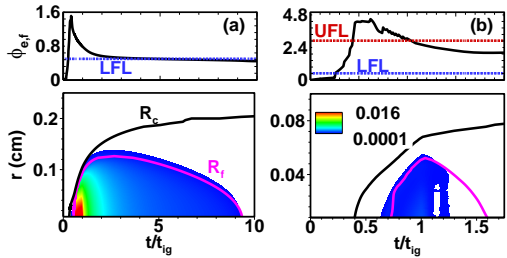

Fig. 6. Time history for $\phi_{e,f}$, $r$-$t$ diagram of FF / ECF trajectories and OH mass fraction for $\phi_{l,0}$ = (a) 1.75 and (b) 6.0. $d_0$ = 5 μm and $E_{ig}$ = 1.0 mJ.

In Fig. 6(a), OH has large mass fraction in the sparking period, up to 0.016. This indicates that the chemical reaction initiated by the spark is strong and thermal runaway proceeds in the gas phase. It is qualitatively consistent with observations in the experiments [8] and DNS [5]. Meanwhile, the flame ER $\phi_{e,f}$ reaches the maximum value of about 1.6 (92% $\phi_{l,0}$), due to continuous droplet evaporation. After the sparking time ( $t > t_{ig}$ ), the foregoing favorable conditions (including the intermediate $\phi_{e,f}$ and on-going thermal runaway) support the kernel to short expanding. It reaches the maximum $R_f$ (0.125 cm) at $2t_{ig}$. Note in passing that, in this period, the flame is in HM state with a negligible distance between FF and ECF. However, the flame ER decreases very quickly after the spark is removed, and hence the chemical reaction is gradually weakened, featured by decreased OH mass fraction. Meanwhile, it is shown from Fig. 6(a) that the ECF still moves outwardly due to the residual heat in the gas, resulting in an increased front distance. This further slows down ethanol vapor diffusion towards the FF. One can see that beyond $t = 4.0t_{ig}$, the flame ER becomes low, only near the LFL. Ultimately, the kernel is quenched at $t = 9.3t_{ig}$.

For $\phi_{l,0}$=6.0 with the same spark energy in Fig. 7(b), the flame kernel is generally much weaker than that in Fig. 7(a). The OH mass fraction is less than 0.001 in the kernel. This means that limited reactions are excited by the spark. Due to heavily loaded liquid ethanol ($\phi_{l,0}$ = 6.0), the flame ER $\phi_{e,f}$ increases to 4.7 before $t < 0.7t_{ig}$, which is too rich to initiate the flame. Meanwhile, the evaporative cooling is strong, which also prevents the thermal runaway of the vapor/air mixture in the spark period. Hence, the weak kernel is generated relatively late, when the flame ER starts to decrease Once the spark is off, it decays instantly in a relatively rich gas (near the UFL of 2.9). This is also observed by de Oliveira [8]. For this scenario, the flame kernel is always located in HM state after the flame kernel is initiated.

Based on our parametric simulations, long failure normally occurs when $\phi_{l,0}$ < ER$_o$. A strong flame kernel can be achieved when the ignition energy is capable to vaporize the droplets around the spark and excite the chemical reaction subsequently. In the long mode, the concentration of fuel vapor in the spark region is above unity, but still below the UFL. It fails due to lean instantaneous flame ER as the kernel expands outwardly. Conversely, short failure mode occurs when $\phi_{l,0}$ > ER$_o$. In this case, the transient flame ER $\phi_{e,f}$ can be much higher than the UFL, particularly for the sparking period. The kernel is born late and weak because of relatively rich mixture and strong droplet heat absorption, and decays fast once the spark is terminated.

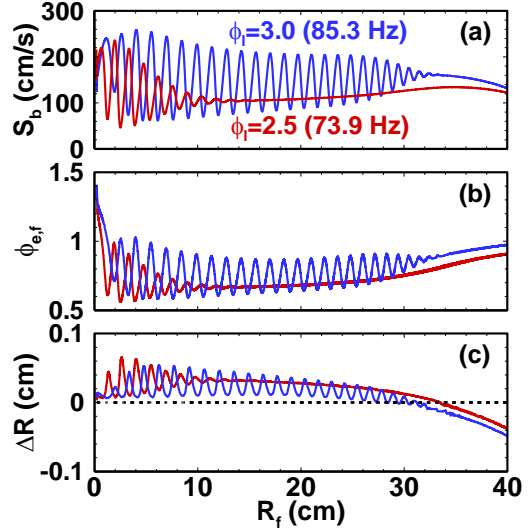

Fig. 7. Time history of (a) flame propagation speed, (b) $\phi_{e,f}$, and (c) $\Delta R$ with $\phi_{l,0}$ = 2.5 and 3.0. $E_{ig}$ = 6.0 mJ, $d_0$ = 10 μm.

### 3.3 Oscillating flame propagation

Up to this point, only early stage of ethanol spray flame initiation is studied. In this section, unsteady flame propagation behaviors in fuel sprays will be discussed. To this end, the computational domain is increased to $r_W$ = 50 cm. Based on our simulations, flame speed oscillation is observed for $d_0 \geq 10 \ \mu m$, and this is because finer droplets approximately follow the gas phase [13,15]. Figure 8 presents the evolutions of flame propagation speed ($S_b$), flame ER and front distance. Here we consider $d_0 = 10 \ \mu m$, but two liquid ERs of $\phi_{l,0} = 2.5$ and 3.0. Evident from Fig. 7 is the regular flame oscillation for both mixtures, excited from the sparking period. The $\phi_{l,0} = 3.0$ case exhibits longer oscillation, which is damped when the flame radius is about 32 cm and the average frequency is about $f = 85.3$ Hz. Here the average frequency is calculated from the five cycles which follow the maximum oscillation amplitude for $S_b$. Nonetheless, for the $\phi_{l,0} = 2.5$ mixture, the flame oscillation terminates earlier (~12 cm), with a lower frequency of 73.9 Hz. These frequencies are generally close to those measured by Sulaiman for *iso*-octane spray spherical flames [15]. It is also seen from Figs. 7(b) and 7(c) that the flame ER oscillates with the same frequency. However, for the front distance, a phase difference around $\pi$ can be found.

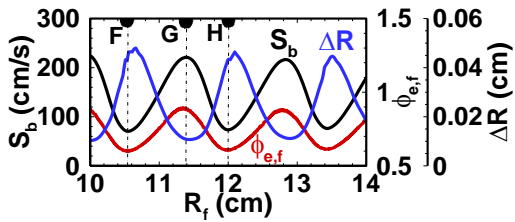

Fig. 8. Time history of flame propagation speed, flame ER, and front distance for $\phi_{l,0} = 3.0$.

An enlarged view of $S_b$, $\phi_{e,f}$, and $\Delta R$ in the selected cycles for $R_f$ from 10 to 14 cm are presented in Fig. 8 when $\phi_{l,0} = 3.0$. At point F, the gas and droplet velocities in the evaporation zone ahead of the flame are approximately equal. The flame accelerates after F and the front distance $\Delta R$ decreases. Due to inertia difference, the gas velocity exceeds the droplet one. The number density of the fuel droplet in the evaporation zone is therefore increased, leading to increased flame ER from F to G, as displayed in Fig. 8. The flame continuously accelerates due to increased fuel vapor availability. At point G, the peak flame ER is reached, which corresponds to the largest flame speed. Hence, the flame decelerates. The droplet velocity shows a slower response and hence is slightly larger than gas velocity. The number density of fuel droplets ahead of the flame decreases, and accordingly the flame ER is reduced. The flame deceleration process ends at H, where the gas and droplet velocities are equal again. The cycle repeats after H. The detailed evolutions of gas velocity, droplet velocity, droplet number density, and chemical heat release for different instants between F and H are presented in the Supplementary Material (Figs. S2 and S3). Since we consider fine droplets, the oscillating flame is always homogeneous. This is different from [13,15], in which periodic shift between HT and HM spray flames is observed when the flame oscillates.

Besides, the number density of liquid droplets in the pre-flame zone decreases when the flame propagates (see supplemental material, Fig. S4) due to the spherical geometry effects. Therefore, the repeated influences between flow and fuel concentration fields cannot maintain. This is evident from the decreased average $\phi_{e,f}$ for each oscillation cycle shown in Fig. 7(b). Besides, for the $\phi_{l,0} = 3.0$ mixture, the flame oscillation damping is also affected by the compression effect from the chamber wall at $r = r_W$. This is shown by the increased pressure at the flame, confirmed by the Figs. S5 and S6 in the supplemental material.

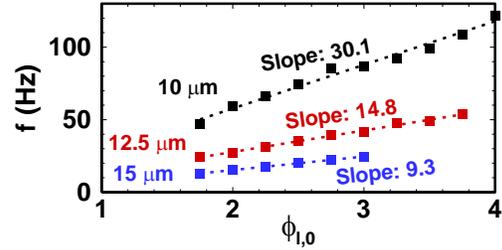

Fig. 9. Flame oscillation frequency versus liquid ER and droplet size.

Dependence of average oscillation frequency on liquid ER and initial droplet size is summarized in Fig. 9. The results in Fig. 9 are achieved with the ignition energy of 6 mJ and it is shown that the predicted frequency is not sensitive to the ignition energy (see a comparison in supplemental material, Fig. S7). For each $d_0$, the frequency linearly increases with the liquid ER. This is because higher liquid ER generally corresponds to larger fuel vapor concentration in the evaporation zone and therefore the variation of the flame ER is stronger for higher $\phi_{l,0}$. Meanwhile, the dependence of the frequency on $\phi_{l,0}$ is gradually reduced when the droplet size increases. Besides, the flame oscillation frequency is smaller for larger droplets. This may be because they have slower evaporation rate and/or longer relaxation timescale.

## 4. Conclusion

The present study numerically investigates the forced ignition and oscillating propagation of spray flame in ethanol droplets/air mixture with a detailed chemical mechanism. Different droplet diameters and liquid equivalence ratios are considered. The concept of evaporation completion front is introduced to study the interaction between the droplet evaporation zone and flame front. The gas composition at the flame front is quantified with the flame ER.

The kernel trajectory is considerably affected by droplet size and liquid ER. Generally, the flame ER reaches the maximum when the evaporation completion front start to evolve from the spherical center. It gradually decreases and reaches a constant value when the flame freely propagates. Quasi-stationary spherical flame is observed when the liquid ER is low, whilst kernel extinction / re-ignition appears when the liquid ER is high. These unsteady flame behaviors are essentially affected by the heat and species diffusion between the droplet evaporation zone and flame front.

The dependence of the minimum ignition energy on liquid ER is U-shaped, with an optimal liquid equivalence ratio range ($ER_o$) with smallest MIEs. Long and short ignition failure modes are observed, respectively for small and large liquid ERs. Below $ER_o$ (long mode), larger ignition energy is required to initiate an expanding kernel, due to the low flame ER under LFL. For ER larger than $ER_o$ (short mode), ignition failure is caused by strong evaporative heat loss and rich gas composition due to heavy droplet loading.

Flame speed oscillation is found when initial droplet size is above 10 $\mu$m from our simulations. The oscillation frequency linearly increases with the liquid ER. Meanwhile, larger droplets correspond to smaller frequencies.

## Acknowledgements

This work is supported by MOE Tier 1 Grant (R-265-000-653-114).